\documentclass[preprint]{revtex4}
\usepackage{graphicx}
\usepackage{dcolumn}
\usepackage{bm}
\newcommand{\non}{\nonumber}
\newcommand{\del}{\partial}
\def\gtsim{\mathrel{\hbox{\raise0.2ex
\hbox{$>$}\kern-0.75em\raise-0.9ex\hbox{$\sim$}}}}
\def\ltsim{\mathrel{\hbox{\raise0.2ex
\hbox{$<$}\kern-0.75em\raise-0.9ex\hbox{$\sim$}}}}

\begin{document}


\title{Electroweak phase transition, critical bubbles 
and sphaleron decoupling condition in the MSSM}

\author{Koichi Funakubo$^{1}$}
\email{funakubo@cc.saga-u.ac.jp}
\author{Eibun Senaha$^2$}%
 \email{senaha@ncu.edu.tw}
\affiliation{$^1$Department of Physics, Saga University,
Saga 840-8502 Japan}
\affiliation{$^2$Department of Physics and Center for Mathematics and Theoretical Physics, 
National Central University, 300 Jhongda Rd. Jhongli, Taiwan 320, R.O.C.}
\bigskip

\date{\today}

\begin{abstract}
The electroweak phase transition and the sphaleron decoupling condition in the 
minimal supersymmetric standard model are revisited
taking the latest experimental data into account.
The light Higgs boson scenario and the ordinary decoupling limit which are classified by the relative size 
between the $CP$-odd Higgs boson mass and $Z$ boson mass 
are considered within the context of electroweak baryogenesis.
We investigate $v/T$ at not only the critical temperature at which the effective potential has two
degenerate minima but also the nucleation temperature of the critical bubbles, where
$v$ is a vacuum expectation value of the Higgs boson and $T$ denotes a temperature.
It is found that $v/T$ at the nucleation temperature 
can be enhanced by about 10\% compared to that at the critical temperature.

We also evaluate the sphaleron decoupling condition including the zero mode factors of the fluctuations
around sphaleron.
It is observed that the sphaleron decoupling condition at the nucleation temperature
is given by $v/T\gtsim 1.38$ for the typical parameter sets. 
In any phenomenologically allowed region, $v/T$ at both the critical and nucleation temperatures  
cannot be large enough to satisfy such a sphaleron decoupling condition.
\end{abstract}

\pacs{Valid PACS appear here}

\maketitle


\section{Introduction}
The origin of the baryon asymmetry of the Universe (BAU) has been a great puzzle 
in particle physics and cosmology. 
From the latest observations of the cosmic microwave background radiations,
the supernovae and the large scale structures, 
the baryon-to-photon ratio is given by~\cite{Amsler:2008zzb}
\begin{eqnarray}
\frac{n_B}{n_\gamma}=(4.7-6.5)\times 10^{-10}\quad (95\%~{\rm C.L.}).
\end{eqnarray}
The nonzero BAU can be generated if the following three conditions are fulfilled~\cite{Sakharov:1967dj}:
(I) baryon number ($B$) violation, (II) $C$ and $CP$ violation,
and (III) departure from thermal equilibrium. 
Electroweak baryogenesis is a scenario in which the BAU is explained based only on 
electroweak physics~\cite{ewbg}.
The standard model (SM) can in principle accommodate above three conditions. However,
it turns out that the SM fails to generate the BAU 
due to the lack of sufficient $CP$ violation~\cite{ewbg_sm_cp},
and the electroweak phase transition (EWPT) is a smooth cross over for $m_h\gtsim73$ GeV~\cite{sm_ewpt}, 
which makes the condition (III) infeasible. 
This failure motivates us to go beyond the SM.
So far, many attempts of electroweak baryogenesis have been made in the various new physics 
models such as the minimal supersymmetric standard model (MSSM)~\cite{ewbg-mssm,Funakubo:2002yb,Carena:2008vj},
the two-Higgs-doublet model~\cite{ewbg-2hdm}, the singlet-extended MSSMs~\cite{ewbg-xMSSM},
and recently the fourth-generation models also have been paid much attention~\cite{ewbg-4sm}.
Among them, the MSSM is a theoretically well-motivated model
and a good candidate for successful electroweak baryogenesis. 

As far as the strong first order EWPT which is necessary for condition (III) is concerned, 
the light Higgs boson is generically favored. 
On the other hand, the negative results of the Higgs boson(s) searches at LEP 
put the constraints on the Higgs boson masses in the MSSM~\cite{Barate:2003sz,Schael:2006cr}.
The analysis has been done based on several benchmarks rather than scanning whole parameter 
space. 
Recently, it is pointed out that the LEP constraints can be relaxed in the more general case.
In Ref.~\cite{Belyaev:2006rf}, it is found that the lightest Higgs boson mass can be smaller than
the $Z$ boson mass.
Such a light Higgs boson can be viable since its coupling to $Z$ boson is modified by the
additional Higgs bosons and suppressed enough to avoid the experimental observed limits.
We call a light Higgs boson scenario (LHS) when the mass of the lightest Higgs boson 
is less than 114.4 GeV. This scenario can accommodate the $2.3\sigma$ excess 
around 98 GeV Higgs boson mass in the LEP experiments~\cite{Barate:2003sz,Schael:2006cr}
and have parameter space that is consistent with the observed dark matter relic density~\cite{DM-LHS}. 
In Ref.~\cite{Kim:2006mb}, it is claimed that the LHS can provide a solution 
for a little hierarchy problem as well.
As for the electroweak baryogenesis context, 
since the neutral Higgs bosons are relatively light, 
they may give a new window to a viable baryogenesis scenario.

In the electroweak baryogenesis mechanism,
a sphaleron process must be decoupled after the EWPT in order to avoid the washout of the generated BAU. 
Because of this requirement, the ratio of the Higgs vacuum expectation value (VEV) 
to the temperature in question 
should be greater than some value.
To search for the region where the strong first order EWPT is realized,
and which is also experimentally allowed,
not only the theoretical calculations of the zero temperature observables 
but also those at the finite temperature are mandatory. 
The accuracy of the former calculations are already beyond at least the tree level.
On the other hand, the sphaleron decoupling condition, which is derived using the tree-level Higgs potential
at zero temperature, has been frequently used to discuss the strength of the strong first order EWPT 
in the literature.
Namely, $v_C>T_C$ is adopted as a practical criterion, where $T_C$ is the critical temperature 
at which the Higgs potential has two degenerate minima and $v_C$ is the Higgs VEV at $T_C$. 
To obtain more accurate theoretical predictions, 
the refined calculation of the sphaleron decoupling condition is indispensable. 
The uncertainties of the condition is expected to be reduced up to about 10\% level
if we make the following improvements, i.e.,
the sphaleron solution at finite temperature is used and 
the contributions from the zero modes of the fluctuations around sphaleron at finite temperature 
are taken into account.
In addition, the sphaleron decoupling condition must be imposed 
at the temperature $T_E$ lower than $T_C$, at which the EWPT terminates.
In practice,  since it is difficult to determine this temperature,
we substitute $T_N$, a nucleation temperature of the critical bubble, in place of $T_E$.
The critical bubble is defined as the bubble whose surface energy and volume energy becomes balanced.
Only such bubbles can nucleate and expand in the symmetric phase.

In this article, the possibility of the strong first order EWPT in the LHS and 
the conventional decoupling limit is examined with a particular emphasis on the 
refinement of the sphaleron decoupling condition.
The analysis is conducted taking the latest experimental data into account as well.
It is well known that a light stop whose mass is smaller than that of top quark 
is required for the strong first order EWPT.
We search for the parameter space that is consistent with such a light stop.
In the LHS, since the masses of all the Higgs bosons are relatively light, typically, $<140$ GeV, 
the $B$ physics observables data can restrict the allowed region.
Here, we consider the constraints from the following processes:
$B_u\to \tau\nu$, $\overline{B}\to X_s\gamma$ and $B_s\to \mu^+\mu^-$.
The first two modes are relevant when the charged Higgs bosons are light. 
The last one, which has not been observed yet, can give a significant limit 
when the neutral Higgs boson is light.

After finding the phenomenologically allowed region, 
we investigate the strength of the EWPT at not only $T_C$ but also $T_N$.
By using the one-loop effective potential at zero and finite temperatures, 
we calculate the sphaleron energy and the contributions from zero modes 
of the fluctuations around sphaleron to evaluate the sphaleron decoupling condition.
It is then demonstrated whether or not such an improved sphaleron decoupling condition is satisfied
at $T_C$ and $T_N$.

This article is organized as follows: In Sec.~\ref{sec:GH}, we present the Lagrangian of the gauge-Higgs
system for describing the statics and dynamics of the EWPT.
The LHS is discussed in Sec.~\ref{sec:LHS}. The qualitative evaluation of the EWPT 
is outlined in Sec.~\ref{sec:EWPT}. The critical bubbles and the sphaleron decoupling condition 
are analyzed in Secs.~\ref{sec:cri_bub} and \ref{sec:sph_dec}, respectively.
The experimental constraints are considered in Sec.~\ref{sec:ex_const}.
In Sec.~\ref{sec:numerical}, we show the numerical results.
Section~\ref{sec:con_dis} is devoted to the conclusion and discussion.
In the Appendix~\ref{app:mass_squared}, we give the mass-squared matrices of the squarks 
to make our notation clear.

\section{Gauge-Higgs system}\label{sec:GH}
To discuss the statics and dynamics of the EWPT, we consider the gauge-Higgs system, which 
is governed by the Lagrangian
\begin{eqnarray}
\mathcal{L}_{\rm gauge-Higgs} =-\frac{1}{4}F^{a}_{\mu\nu}F^{a\mu\nu}
	-\frac{1}{4}B_{\mu\nu}B^{\mu\nu}+ \sum_{i=d, u}(D_\mu\Phi_i)^\dagger D^\mu\Phi_i
	-V_{\rm eff}(\Phi_d, \Phi_u; T).
\label{gauge-Higgs}
\end{eqnarray}
The covariant derivatives are
\begin{eqnarray}
D_\mu\Phi_d = \left(\del_\mu+ig_2\frac{\tau^a}{2}A^a_\mu-i\frac{g_1}{2}B_\mu\right)\Phi_d,\quad
D_\mu\Phi_u = \left(\del_\mu+ig_2\frac{\tau^a}{2}A^a_\mu+i\frac{g_1}{2}B_\mu\right)\Phi_u,
\end{eqnarray}
where $\tau^a$ is the Pauli matrices, 
$g_2$ and $g_1$ are the gauge couplings for $SU(2)_L$ and $U(1)_Y$, respectively.
$V_{\rm eff}$ is the effective potential, which is composed of the tree and one-loop parts
\begin{eqnarray}
V_{\rm eff}(\Phi_d, \Phi_u; T)=V_0(\Phi_d, \Phi_u)+\Delta V(\Phi_d, \Phi_u; T).
\end{eqnarray}
The tree-level part is given by the F-, D-, and the soft supersymmetry (SUSY) breaking terms
\begin{eqnarray}
V_0(\Phi_d, \Phi_u) &=& m_{1}^{2}\Phi_{d}^\dagger\Phi_d
	+m_{2}^{2}\Phi_{u}^\dagger\Phi_u
	-(m_{3}^{2}\epsilon_{ij}\Phi^i_{d}\Phi^j_{u}+\mbox{h.c.}) \non\\
&&+\frac{g^2_2+g^2_1}{8}(\Phi^\dagger_{d}\Phi_d-\Phi^\dagger_{u}\Phi_u)^{2}
	+\frac{g^2_2}{2}(\Phi^\dagger_d\Phi_u)(\Phi^\dagger_u\Phi_d),
\end{eqnarray}
where
\begin{eqnarray}
m^2_1 = \tilde{m}^2_1+|\mu|^2,\quad m^2_2 = \tilde{m}^2_2+|\mu|^2,\quad m^2_3 = \mu B.
\end{eqnarray}
The Higgs doublets are parameterized as
\begin{eqnarray}
\Phi_d&=&
	\left(
		\begin{array}{c}
			\frac{1}{\sqrt{2}}(v_d+h_d+ia_d) \\
			\phi_d^-
		\end{array}
	\right),\quad 
\Phi_u=
	e^{i\vartheta}\left(
		\begin{array}{c}
			\phi_u^+\\
			\frac{1}{\sqrt{2}}(v_u+h_u+ia_u) 
		\end{array}
	\right),
\end{eqnarray}
where we assume that $U(1)_{\rm em}$ is not broken at the vacuum.
Unlike the general two-Higgs-doublet model~\cite{HHG}, $CP$ violation cannot be accommodated 
in the Higgs potential at the tree level, leading to $\vartheta=0$.

The one-loop part comprises the effective potentials at zero and nonzero temperatures
\begin{eqnarray}
\Delta V(\Phi_d, \Phi_u; T)=\sum_A c_A\left[F_0(\bar{m}^2_A)
	+\frac{T^4}{2\pi^2}I_{B, F}\left(\frac{\bar{m}^2_A}{T^2}\right)\right],
\end{eqnarray}
where
\begin{eqnarray}
F_0(m^2)=\frac{m^4}{64\pi^2}\left(\ln\frac{m^2}{M^2}-\frac{3}{2}\right),\quad
I_{B, F}(a^2)=\int^\infty_0 dx~x^2\ln\left(1\mp e^{-\sqrt{x^2+a^2}}\right).
\end{eqnarray}
where $F_0$ is the zero temperature effective potential, which is regularized 
in the $\overline{\rm DR}$ scheme, 
$c_A$ denotes the degrees of freedom of particle species $A$, 
$\bar{m}_A$ is the background-field-dependent mass, and
$M$ is a renormalization scale, which is determined in such a way that 
the zero temperature one-loop correction to the effective potential vanishes in the vacuum.
In what follows, the contributions of the weak gauge bosons ($Z, W$), 
the third-generation quarks ($t, b$), and squarks ($\tilde{t}_{1,2},\tilde{b}_{1,2}$) are taken into account.
The statistical factors of each particle is, respectively, given by 
\begin{eqnarray}
c_Z = 3,\quad c_W = 6,\quad c_t=c_b=-4N_c,\quad c_{\tilde{t}_{1,2}}=c_{\tilde{b}_{1,2}}=2N_c,
\end{eqnarray}
where $N_c$ is the number of color.
In this article, we use the one-loop mass formulae of the Higgs bosons presented 
in Ref.~\cite{Funakubo:2002yb}.

\section{Light Higgs boson scenario}\label{sec:LHS}
To make a discussion on the LHS clear, we rotate the two Higgs doublets $(i\tau^2\Phi^*_d,\Phi_u)$ 
into $(i\tau^2\Phi^{'*}_d,\Phi'_u)$ by $\beta=\tan^{-1}(v_u/v_d)$. 
For simplicity, we assume that $CP$ is conserved for the moment.
The rotated Higgs doublets are cast into the form
\begin{eqnarray}
\Phi'_d=
	\left(
	\begin{array}{c}
		\frac{1}{\sqrt{2}}(v_0+h'_d+iG^0)\\
		G^-
	\end{array}
\right),\quad
\Phi'_u=
\left(
	\begin{array}{c}
		H^+\\
		\frac{1}{\sqrt{2}}(h'_u+iA)
	\end{array}
\right),\label{rotated_phi}
\end{eqnarray}
where $v_0=\sqrt{v^2_{d}+v^2_{u}}\simeq 246$ GeV,
 $A$ is the $CP$-odd Higgs boson, $H^+$ is the charged Higgs boson
and ($G^0, G^\pm$) are the Nambu-Goldstone bosons.
$h'_d$ and $h'_u$ are related to the physical $CP$-even Higgs bosons ($h, H$) via
\begin{eqnarray}
	\left(
		\begin{array}{c}
			h'_d \\
			h'_u
		\end{array}
	\right)
=
	\left(
		\begin{array}{cc}
			\cos(\beta-\alpha) & \sin(\beta-\alpha) \\
			-\sin(\beta-\alpha) & \cos(\beta-\alpha)
		\end{array}
	\right)	
	\left(
		\begin{array}{c}
			H \\
			h
		\end{array}
	\right),\label{rotated_h}
\end{eqnarray}
where $\alpha$ is a mixing angle between $h_d$ and $h_u$.
It should be noticed that only $h'_d$ possesses the Higgs VEV in the rotated basis, 
and we can identify it the SM-like Higgs boson, i.e., $h'_d=h^{\rm SM}$.

At the tree level, the Higgs boson couplings to $Z$ boson normalized 
to the corresponding SM values are, respectively, given by
\begin{eqnarray}
&&\frac{g_{hZZ}}{g_{h^{\rm SM}ZZ}} 
	= \sin(\beta-\alpha),\quad 
	\frac{g_{HZZ}}{g_{h^{\rm SM}ZZ}} = \cos(\beta-\alpha),\label{Higgs_ZZ}\\
&&\frac{g_{hZA}}{g_{h^{\rm SM}ZG^0}} = \cos(\beta-\alpha),\quad 
	\frac{g_{HZA}}{g_{h^{\rm SM}ZG^0}} = \sin(\beta-\alpha).\label{Higgs_ZA}
\end{eqnarray}
From Eqs.~(\ref{rotated_phi}), (\ref{rotated_h}), (\ref{Higgs_ZZ}) and (\ref{Higgs_ZA}), 
one can see that $h\to h^{\rm SM}$, $g_{hZZ}\to g_{h^{\rm SM}ZZ}$ and
$g_{hZA}\to 0$ for $\sin(\beta-\alpha)\to 1$
and $H\to h^{\rm SM}$, $g_{HZZ}\to g_{h^{\rm SM}ZZ}$ and
$g_{HZA}\to 0$ for $\cos(\beta-\alpha)\to 1$.
The former case (\lq\lq decoupling limit\rq\rq) is realized by taking $m_Z\ll m_A$ 
and the latter one (\lq\lq antidecoupling limit\rq\rq) is possible for $m_A\sim m_Z$, 
which corresponds to the LHS.
Depending on the theory parameters, some of the above four Higgs boson couplings 
can be small enough to avoid the LEP exclusion limits. 
In the LHS, for example, $h$ can be as light as about 100 GeV
for $g_{hZZ}/g_{h^{\rm SM}ZZ}\simeq 0.5$.
However, from the sum rules among the Higgs boson couplings, i.e., 
$(g^2_{hZZ}+g^2_{HZZ})/g^2_{h^{\rm SM}ZZ}=(g^2_{hZA}+g^2_{HZA})/g^2_{h^{\rm SM}ZG^0}=1$,
not all Higgs bosons are allowed to be light. If the Higgs boson is the SM-like,
the mass of it should be larger than 114.4 GeV since its coupling to the $Z$ boson
becomes the SM value. We will discuss the experimental constraints in more detail in Sec.~\ref{sec:ex_const}.

Finally, we comment on the $CP$-violating case.
Although $CP$ is conserved in the tree-level Higgs sector, it can be broken by the radiative corrections
such as the top/stop loops. The realization of the maximal $CP$ violation 
is called the CPX scenario~\cite{CPX,Lee:2007gn}.
In this scenario, it is possible to accommodate an even lighter Higgs boson.
For instance, the mass of the lightest Higgs boson can be as small as 40 GeV, which is experimentally viable.
It is interesting to discuss the phenomenological consequences of this scenario. 
However, the CPX scenario requires the relatively large ratios of
$(\mu/m_{\tilde{q}}, A_t/m_{\tilde{q}})$ assuming $m_{\tilde{q}}=m_{\tilde{t}_R}$,
which is not favored in the context of the strong first order EWPT that we will discuss in the next section.

\section{Electroweak phase transition}\label{sec:EWPT}
To explain how the first order EWPT is strengthened in the MSSM, we here give a brief review on 
the light stop scenario. To simplify the discussion, we use the high temperature expansion
in the effective potential~\cite{Dolan:1973qd}.
In the decoupling limit, $h$ is responsible for the electroweak symmetry breaking 
since $h=h^{\rm SM}$. 
On the other hand, $v$ is mostly shared by $H$ in the LHS.

If the EWPT is first order, the critical temperature $(T_C)$ is given by the temperature at which
the effective potential has two degenerate minima. Let $v_C$ be the nonzero VEV at $T_C$.
From the argument using the high temperature expansion: $a\ltsim1$ in $I_{B,F}(a^2)$, 
one can see that $v_C/T_C$ can be enhanced when the quartic term in $V_{\rm eff}$
gets smaller and/or the coefficient of the cubic term with a negative coefficient 
(denoted $E$) becomes larger.
The former implies that the light $h/H$ is favored in the decoupling/LHS limit.
The latter can be realized by the additional contributions from the bosons.
In fact, the effect of the lighter stop $\tilde{t}_1$ on $E$ is sizable due to the 
large top Yukawa coupling constant ($y_t$) and the degrees of freedom, i.e., $2N_c$.
In the high temperature expansion, one can obtain
\begin{eqnarray}
V_{\rm eff}(v;T) \ni -ETv^3\simeq-(E_{\rm SM}+E_{\tilde{t}_1})Tv^3,
\end{eqnarray}
where 
\begin{eqnarray}
E_{\rm SM}\simeq\frac{1}{4\pi v^3_0}(2m^3_W+m^3_Z)\simeq 0.01,\quad
E_{\tilde{t}_1} \simeq \frac{N_c|y_t|^3\sin^3\beta}{12\sqrt{2}\pi}
\left( 1 - \frac{|X_t|^2}{m_{\tilde q}^2} \right)^{3/2},\label{delE_HTE}
\end{eqnarray}
where $X_t=A^*_t-\mu/\tan\beta$ and the $\mathcal{O}(g^2)$ contributions are neglected
in the stop contribution, also the hierarchies between the soft SUSY breaking masses
are implicitly assumed, namely, $m_{\tilde{q}}\gg X_t, m_{\tilde{t}_R}=0$ are taken
to be consistent with the LEP bounds on the Higgs boson mass and the $\rho$ parameter.
As a result of $m_{\tilde{t}_R}=0$, which is necessary for the enhancement of the loop effect, 
the lighter stop mass becomes less than the top mass $m_{\tilde{t}_1}<m_t$.
From Eq.~(\ref{delE_HTE}), one can easily see that the no-mixing case 
$X_t=0$ maximizes the loop effect.
If we assume $\tan\beta=\tan\beta(T=0)$, we obtain $E_{\tilde{t}_1}\sim 0.056$,
which is about 6 times as large as the SM contributions.

To make the analysis on the EWPT precisely, the corrections from the daisy diagrams 
must be taken into account, which yield the temperature dependent terms in the field-dependent masses
of particles. 
For the light stop, we effectively have the following mass shift to leading order:
\begin{eqnarray}
\hat{m}^2_{\tilde{t}_R}(T)=m^2_{\tilde{t}_R}+cT^2,
\end{eqnarray}
where $c$ is positive and composed of the relevant couplings in the theory. 
Therefore, the EWPT would get weaker even if $m^2_{\tilde{t}_R}=0$.
Nonetheless, as advocated in Ref.~\cite{Carena:1996wj} the choice of $m^2_{\tilde{t}_R}=-cT^2$, 
which can induce the charge-color-breaking (CCB) vacuum,
may still realize the same enhancement of the stop loop effect as outlined above.
For simplicity, we will not pursue this possibility and take $m^2_{\tilde{t}_R}\simeq0$ putting
the daisy diagrams aside in this article.
Our analysis on the strength of the EWPT is expected to be the same, or might be more optimistic 
than the more realistic case depending on the model parameters.

The high temperature expansion makes it easy to see how the first order phase transition is strengthened
analytically. It is, however, untrustworthy to use the approximation 
when the masses of the particles in loops are somewhat larger than $T_C$.  
We thus adopt a different method.
Although the numerical integrations contained in the definitions of $I_{B,F}(a^2)$ are the standard way, 
it is an extremely time-consuming task. 
Therefore, we will use the following fitting functions in our numerical analysis:
\begin{eqnarray}
\tilde{I}_{B,F}(a^2)=e^{-a}\sum^N_{n=0}c^{b,f}_na^n,\label{I_fit}
\end{eqnarray}
where the coefficients $c^{b,f}_n$ are determined by the least squared method.
For $N=40$, the errors of $\tilde{I}_{B,F}(a^2)$ do not exceed $10^{-6}$ for any $a$,
which is sufficient for our purpose.
To determine $T_C$ and $v_C$, we minimize $V_{\rm eff}$ with Eq.~(\ref{I_fit}) in the three dimensional space 
$(v_d, v_u\cos\vartheta, v_u\sin\vartheta)$ numerically.

\section{Critical bubbles}\label{sec:cri_bub}
The first order EWPT begins at somewhat below $T_C$. 
For the EWPT to proceed, the radius of the bubble should be larger than
some critical size, otherwise it would shrink by the surface tension of the bubble wall.
The bubble of this critical size is called the critical bubble.
After nucleation of the critical bubbles, they start to percolate and eventually convert 
the symmetric phase into the broken phase if the supercooling is not too large.

From Eq.~(\ref{gauge-Higgs}), the energy functional in the temporal gauge takes the form
\begin{eqnarray}
E = \int d^3\boldsymbol{x} 
	\bigg[\frac{1}{4}F^a_{ij}F^a_{ij}+\frac{1}{4}B_{ij}B_{ij}
	+(D_i\Phi_d)^\dagger D_i\Phi_d+(D_i\Phi_u)^\dagger D_i\Phi_u
	+V_{\rm eff}(\Phi_d, \Phi_u;T)\bigg].
\end{eqnarray}
Here, we assume that the least energy has the pure-gauge configurations 
for $A^a(x)$ and $B(x)$, hence $F^a_{ij}=B_{ij}=0$.
Since a spherically symmetric configuration can give the least energy, 
the Higgs fields depend only on radial coordinate $r=\sqrt{\boldsymbol{x}^2}$.
The classical Higgs fields are parameterized as
\begin{eqnarray}
\Phi_d(r)=\frac{e^{i\theta_d(r)}}{\sqrt{2}}
\left(
	\begin{array}{c}
		\rho_d(r) \\
		0
	\end{array}
\right),\quad
\Phi_u(r)=\frac{e^{i(\theta_u(r)+\rho)}}{\sqrt{2}}
\left(
	\begin{array}{c}
		0 \\
		\rho_u(r)
	\end{array}
\right),\label{rho_fields}
\end{eqnarray}
where the gauge-invariant constant phase $\rho$ will be taken to make the boundary
value $\theta_d+\theta_u$ at $r=\infty$ to vanish.
To remove the energy source for the $Z$ boson from the Higgs current,
the so-called ``sourcelessness condition"~\cite{cpv_bubble1}
\begin{eqnarray}
	\rho^2_d\frac{d \theta_d}{dr}
	-\rho^2_u\frac{d \theta_u}{dr}=0
\label{sourcelessness}
\end{eqnarray}
must be satisfied. 
Now let us introduce $\theta=\theta_d+\theta_u$ and $\bar{\theta}=\theta_d-\theta_u$.
One can easily see that only $\theta$ is a gauge-independent phase.
Equation~(\ref{sourcelessness}) can be rewritten as
\begin{eqnarray}
	(\rho^2_d+\rho^2_u)\frac{d \bar{\theta}}{dr}
	+(\rho^2_d-\rho^2_u)\frac{d \theta}{dr} = 0,
\end{eqnarray}
which is used to eliminate $\bar{\theta}$.
The energy functional is then reduced to
\begin{eqnarray}
E = 4\pi\int^\infty_0dr~r^2\bigg[\frac{1}{2}\bigg\{\bigg(\frac{d\rho_d}{dr}\bigg)^2
	+\bigg(\frac{d\rho_u}{dr}\bigg)^2\bigg\} 
	+\frac{1}{2}\frac{\rho^2_d\rho^2_u}{\rho^2_d+\rho^2_u}\bigg(\frac{d\theta}{dr}\bigg)^2
	+V_{\rm eff}(\rho_d,\rho_u,\theta;T)\bigg].
\end{eqnarray}
From this, the equations of motion (EOM) for $\rho_d,\rho_u$ and $\theta$ are, respectively, given by
\begin{eqnarray}
-\frac{1}{r^2}\frac{d}{dr}\bigg(r^2\frac{d\rho_d}{dr}\bigg)
	+\rho_d\bigg(\frac{\rho^2_u}{\rho^2_d+\rho^2_u}\frac{d\theta}{dr}\bigg)^2
	+\frac{\del V_{\rm eff}}{\del \rho_d}&=&0, \\
-\frac{1}{r^2}\frac{d}{dr}\bigg(r^2\frac{d\rho_u}{dr}\bigg)
	+\rho_u\bigg(\frac{\rho^2_d}{\rho^2_d+\rho^2_u}\frac{d\theta}{dr}\bigg)^2
	+\frac{\del V_{\rm eff}}{\del \rho_u}&=&0, \\
-\frac{1}{r}\frac{d}{dr}\bigg(r^2\frac{\rho^2_d\rho^2_u}{\rho^2_d+\rho^2_u}\frac{d\theta}{dr}\bigg)
	+\frac{\del V_{\rm eff}}{\del \theta}&=&0.
\end{eqnarray}
The solutions can exist only when the temperature lies in the range $T_0<T<T_C$, 
where $T_0$ is the temperature at which
the Higgs potential at the origin is destabilized in some direction.
Since we focus on the $CP$-conserving bubble walls~\cite{Moreno:1998bq},
$\theta(r)$ is set to be zero in the following discussion. 
The studies of the $CP$-violating bubble walls can be found 
in Refs.~\cite{cpv_bubble1,cpv_bubble2}.
The boundary conditions for EOM are imposed in the symmetric ($r=\infty$) and broken ($r=0$) phases as
\begin{eqnarray}
\lim_{r\to\infty}\rho_d(r)=0,\quad \lim_{r\to\infty}\rho_u(r)=0,\quad
\frac{d\rho_d(r)}{dr}\bigg|_{r=0}=0,\quad\frac{d\rho_u(r)}{dr}\bigg|_{r=0}=0.
\end{eqnarray}
It is convenient to parameterize the Higgs profiles $(\rho_d, \rho_u)$
in terms of the dimensionless quantities
\begin{eqnarray}
\xi = vr,\quad h_1(\xi)=\frac{\rho_d(r)}{v\cos\beta},\quad h_2(\xi)=\frac{\rho_u(r)}{v\sin\beta}.
\end{eqnarray}
Then $E$ takes the form
\begin{eqnarray}
E &=& 4\pi v\int^\infty_0 d\xi~\xi^2\bigg[
	\frac{1}{2}\bigg\{\bigg(\frac{dh_1}{d\xi}\bigg)^2\cos^2\beta
	+\bigg(\frac{dh_2}{d\xi}\bigg)^2\sin^2\beta\bigg\}
	+\tilde{V}_{\rm eff}(h_1,h_2;T)\bigg],
\end{eqnarray}
where $\tilde{V}_{\rm eff}=V_{\rm eff}/v^4$.
Correspondingly, the EOM are rewritten as
\begin{eqnarray}
-\frac{1}{\xi^2}\frac{d}{d\xi}\bigg(\xi^2\frac{dh_1}{d\xi}\bigg)
	+\frac{1}{\cos^2\beta}\frac{\del \tilde{V}_{\rm eff}}{\del h_1}&=&0,\label{EOM_h1}\\
-\frac{1}{\xi^2}\frac{d}{d\xi}\bigg(\xi^2\frac{dh_2}{d\xi}\bigg)
	+\frac{1}{\sin^2\beta}\frac{\del \tilde{V}_{\rm eff}}{\del h_2}&=&0\label{EOM_h2},
\end{eqnarray}
with the boundary conditions
\begin{eqnarray}
\lim_{\xi\to\infty}h_1(\xi)=0,\quad \lim_{\xi\to\infty}h_2(\xi)=0,\quad
\frac{dh_1(\xi)}{d\xi}\bigg|_{\xi=0}=0,\quad\frac{dh_2(\xi)}{d\xi}\bigg|_{\xi=0}=0\label{BC_h}.
\end{eqnarray}
We will comment on the numerical method to solve the EOM for $h_1(\xi)$ and $h_2(\xi)$ 
in Sec. \ref{subsec:relaxation}.
%
%
\subsection{Bubble nucleation temperature}
The bubble nucleation rate per unit time per unit volume is given by~\cite{Linde:1981zj}
\begin{eqnarray}
\Gamma_N(T)\simeq T^4\left(\frac{E_{\rm cb}(T)}{2\pi T}\right)^{3/2}e^{-E_{\rm cb}(T)/T}.
\end{eqnarray}
where $E_{\rm cb}(T)$ is the energy of the critical bubble at temperature $T$. 
Note that this is a rate per unit volume. We define the nucleation temperature $T_N$ 
as the temperature at which the rate of the nucleation of a critical bubble within 
a horizon volume is equal to the Hubble parameter at that temperature. 
Since the horizon scale is roughly given by $H^{-1}(T)$, the nucleation temperature is defined by
\begin{eqnarray}
\Gamma_N(T_N)/H^3(T_N)=H(T_N)\simeq 1.66\sqrt{g_*(T_N)}T^2_N/m_{\rm P},
\label{def_Tn}
\end{eqnarray}
where $g_*(T_N)$ is the massless degrees of freedom at $T_N$ 
and $m_{\rm P}$ is the Planck mass ($\simeq 1.22\times 10^{19}$ GeV).
Since a single bubble nucleated within the horizon volume 
cannot convert the whole region to the broken phase, the nucleation temperature 
defined by Eq.~(\ref{def_Tn}) simply gives an upper bound of the temperature at which
the EWPT starts.
To determine $T_N$ more precisely, simulations or the methods employed
in Ref.~\cite{Carrington:1993ng} are needed.

From Eq.~(\ref{def_Tn}), it follows that
\begin{eqnarray}
\frac{E_{\rm cb}(T_N)}{T_N}-\frac{3}{2}\ln\left(\frac{E_{\rm cb}(T_N)}{T_N}\right) = 
152.59-2\ln g_*(T_N)-4\ln\left(\frac{T_N}{100~{\rm GeV}}\right).
\end{eqnarray}
Roughly, $E_{\rm cb}/T\ltsim150$ is necessary for the development of the EWPT.
%
%
\subsection{Numerical algorithm for critical bubbles}\label{subsec:relaxation}
To solve the EOM (\ref{EOM_h1}) and (\ref{EOM_h2}) 
under the boundary conditions (\ref{BC_h}), we use ``relaxation methods."
To implement the relaxation methods successfully, the initial configurations of the bubbles 
are of great importance. Here, we take the kink-type ansatz for them, i.e., 
\begin{eqnarray}
h_1(\xi)=h_2(\xi) = \frac{1-\tanh\big\{(\xi-R)/L_w\big\}}{1-\tanh(-R/L_w)},
\end{eqnarray}
where $R$ is the radius of the bubbles and $L_w$ is the bubble wall width.
Since the best values of $R$ and $L_w$ can vary as the temperature goes down, 
it is impossible to know them in advance.
Nonetheless, it turns out that in most cases $10\ltsim L_w<R\ltsim100$ 
can give the successful initial configurations as long as $T$ lies somewhere between $T_N$ and $T_C$.
After obtaining the convergent solutions at some temperature, 
we use them as the initial configurations to search for the critical bubbles at the temperature which is near 
the previous one. By doing this process iteratively, we find $T_N$.

\section{Sphaleron decoupling condition}\label{sec:sph_dec}
As shown by Manton in Ref.~\cite{Manton:1983nd}, the static and unstable classical solution with a finite energy 
can exist in the $SU(2)$ gauge theory.
Such a configuration is called sphaleron, which has a maximal energy along the least energy path.
Similar to the instanton process at zero temperature, the baryon number can be violated 
by the sphaleron process at finite temperature.
Although the instanton process, which is quantum tunneling, is highly suppressed and unobservable, 
the sphaleron process, which is thermal fluctuation, can be active at high temperature, especially 
before the electroweak symmetry is broken.

In order to avoid the washout of the generated BAU after the phase transition,
the sphaleron process must be decoupled. This condition can be obtained by
demanding that the baryon number changing rate 
be smaller than the Hubble parameter~\cite{Arnold:1987mh}
\begin{eqnarray}
-\frac{1}{B}\frac{dB}{dt}\simeq \frac{13N_f}{4\cdot 32\pi^2}\frac{\omega_-}{\alpha^3_W}\kappa\mathcal{N}_{\rm tr}\mathcal{N}_{\rm rot}e^{-E_{\rm sph}/T}<H(T),
\label{sph_dec1}
\end{eqnarray}
where $N_f$ is number of generation, $\alpha_W=\alpha_{\rm em}/\sin^2\theta(m_Z)$, 
$\omega_-$ is the negative mode of the fluctuations around sphaleron,
$\kappa$ is the $\mathcal{O}(1)$ coefficient~\cite{Carson:1990jm,Moore:1998swa}, 
$\mathcal{N}_{\rm tr}$ and $\mathcal{N}_{\rm rot}$ 
are contributions from the translational and rotational zero modes, respectively, 
$E_{\rm sph}$ is the sphaleron energy.
If we denote the sphaleron energy as $E_{\rm sph}=4\pi v\mathcal{E}/g_2$, 
Eq.~(\ref{sph_dec1}) can be translated into 
\begin{eqnarray}
\frac{v}{T}&>&\frac{g_2}{4\pi\mathcal{E}}
	\bigg[42.97+\ln(\kappa\mathcal{N}_{\rm tr}\mathcal{N}_{\rm rot})
		+\ln\bigg(\frac{\omega_-}{m_W}\bigg)
		-\frac{1}{2}\ln\bigg(\frac{g_*}{106.75}\bigg)
		-2\ln\bigg(\frac{T}{100~{\rm GeV}}\bigg)
	\bigg]. \label{sph_cond}
\end{eqnarray}
It should be noted that model dependent parameters other than the sphaleron energy 
in the right-hand side of Eq.~(\ref{sph_cond}) contribute only logarithmically.

In the SM~\cite{Esph_footnote}, the sphaleron solution obtained with the tree-level Higgs
potential yields $\mathcal{E}=2.00$~\cite{Klinkhamer:1984di}, 
$\mathcal{N}_{\rm tr}\mathcal{N}_{\rm rot}=80.13$~\cite{Carson:1989rf,Carson:1990jm},
$\omega^2_-=2.3m^2_W$~\cite{Akiba:1989xu,Carson:1989rf,Carson:1990jm}
for $\lambda/g_2^2=1$. Inserting these values into Eq.~(\ref{sph_cond}) together with
$\kappa=1$,~$g_*=106.75$, and $T=100$ GeV, we find
\begin{eqnarray}
\frac{v}{T}>0.026\times(42.97+4.38+0.416)=1.24,
\end{eqnarray}
where $g_2=2\sqrt{\pi\alpha_{\rm em}(m_Z)}/\sin\theta_W(m_Z)=0.652$~\cite{Amsler:2008zzb} is used.
It is found that the contributions of the zero mode factors to $v/T$ can reach around 10\%,
while the contributions from the last three terms in the right-hand side of Eq.~(\ref{sph_cond})
are only about 1\%.

\begin{center}
\begin{figure}[t]
\includegraphics[width=6.7cm]{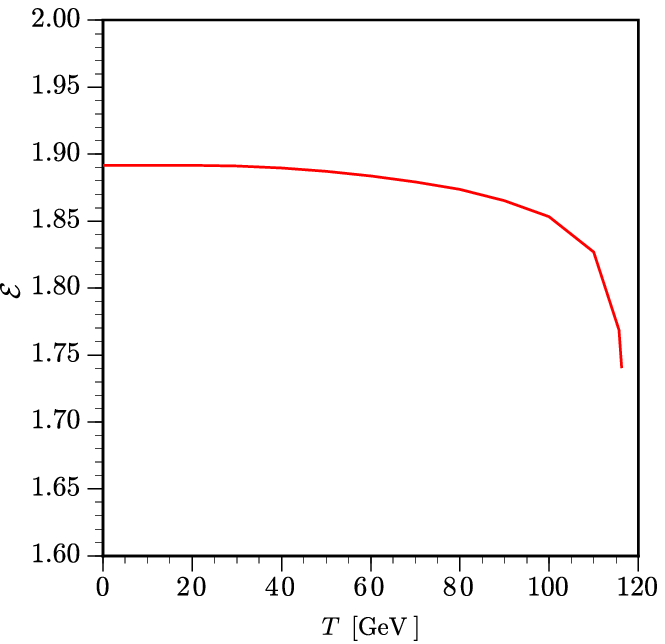}
\hspace{1cm}
\includegraphics[width=6.5cm]{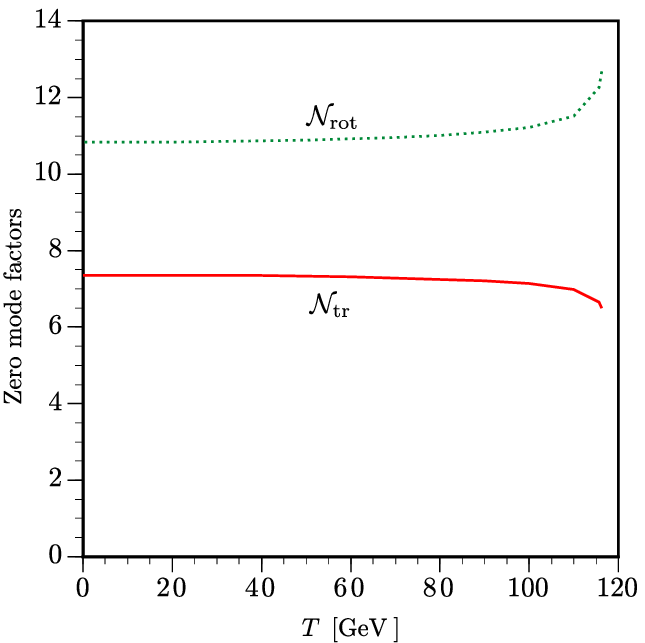}
\caption{The temperature dependences of the sphaleron energy $\mathcal{E}=E_{\rm sph}/(4\pi v/g_2)$ 
and the zero mode factors $\mathcal{N}_{\rm tr, rot}$.}
\label{fig:sph_profile}
\end{figure}
\end{center}

Now we discuss the sphaleron decoupling condition in the MSSM.
As an example, we take $\tan\beta=10.11$, $m_{H^\pm}=127.4$ GeV, $A_t=A_b=-300$ GeV, 
$\mu=100$ GeV, $m_{\tilde{q}}=1200$ GeV, $m_{\tilde{t}_R}=10^{-4}$ GeV.
This parameter set is experimentally allowed as we will discuss in Sec.~\ref{sec:numerical}.
To see the effects of the temperature dependence of the sphaleron solution and the zero mode
factors on the sphaleron decoupling condition, 
we consider the following three cases:
\begin{itemize}
\item[I:] The sphaleron decoupling condition without the zero mode factors 
based on the zero temperature potential $V_{\rm eff}(T=0)$.
\item[II:] The sphaleron decoupling condition with the zero mode factors 
based on the zero temperature potential $V_{\rm eff}(T=0)$.
\item[III:] The sphaleron decoupling condition with the zero mode factors 
based on the full effective potential $V_{\rm eff}(T\neq0)$.
\end{itemize}
The result of each case is summarized in Table \ref{tab:sph_cond}.
$T=T_N$ is used in the right-hand side of Eq.~(\ref{sph_cond}).
As we see from I and II, the contribution of the zero mode factors to 
the sphaleron decoupling condition can be as large as 10\% level, which is the same as in the SM case.
In case III, $v_N/T_N>1.38$ is obtained, which is about 40\% stronger 
than the usual rough estimate used in the literature. 
We look into this case in more detail in the following.
The temperature dependences of $\mathcal{E}$, $\mathcal{N}_{\rm tr}$ and $\mathcal{N}_{\rm rot}$
are plotted in Fig.~\ref{fig:sph_profile}. 
The endpoints of those curves correspond to those at $T_C$.
The left panel shows that the sphaleron energy decreases as $T$ increases.
The change of the sphaleron energy is sizable around $T_C$, hence
which can be the dominant error of the sphaleron decoupling condition if we neglect this effect.
On the other hand, as shown in the right panel the temperature dependences of $\mathcal{N}_{\rm tr}$ 
and $\mathcal{N}_{\rm rot}$ are mild and thus have little effects on $v/T$.
From this analysis, we conclude that 
in order to evaluate the sphaleron decoupling condition within a 10\% accuracy,
the sphaleron solution and the zero mode factors at finite temperature must be taken into account.

Here, we comment on the earlier work~\cite{Moreno:1996zm}.
The authors have evaluated sphaleron at finite temperature
using the one-loop finite-temperature effective potential with the daisy resummations.
In their analysis, only the energy was calculated and 
the zero mode factors were missing.
To search for the possible region where the EWPT is strong first order, they used the following 
sphaleron decoupling condition~\cite{Esph-T}:
\begin{eqnarray}
\frac{E_{\rm sph}(T_C)}{T_C}>45.
\end{eqnarray}
In our calculation including the zero mode factors, 
this condition is translated into 
\begin{eqnarray}
\frac{v_N}{T_N}>1.32
\end{eqnarray}
in case III.

\begin{table}[h]
\begin{center}
\begin{tabular}{|c|c|c|c|}
\hline
& I & II & III \\
\hline
$\mathcal{E}$ & 1.89 & 1.89 & 1.77 \\  
$\mathcal{N}_{\rm tr}$ & --- & 7.36 & 6.65 \\
$\mathcal{N}_{\rm rot}$ & --- & 10.84 & 12.27 \\
\hline
$v_N/T_N>$ & 1.17 & 1.29 & 1.38 \\
\hline
\end{tabular}
\end{center}
\caption{The sphaleron decoupling conditions in the three cases I, II, and III.}
\label{tab:sph_cond}
\end{table}

\section{Experimental constraints} \label{sec:ex_const}
In this section, we present the experimental constraints that we will impose in the numerical calculation.
When the masses of the three neutral Higgs boson ($H_i$) are smaller than 114.4 GeV
and/or the sum of two of them are smaller than about 195 GeV, we require
\begin{eqnarray}
&&g^2_{H_iZZ}\times {\rm Br}(H_i\to f\bar{f}) < \mathcal{F}_{H_iZ}(m_{H_i}), \non\\
&&g^2_{H_iH_jZ}\times {\rm Br}(H_i\to f\bar{f})\times {\rm Br}(H_j\to f\bar{f})
 < \mathcal{F}_{H_iH_j}(m_{H_i}+m_{H_j}),
\label{LEP-Higgs}
\end{eqnarray}
where $f=b,\tau$. 
$\mathcal{F}_{H_iZ}$, and $\mathcal{F}_{H_iH_j}$ are 
the 95\% C.L. upper limits from the LEP experiments~\cite{Barate:2003sz,Schael:2006cr}.  
The Higgs boson couplings to the $Z$ boson normalized to their SM values are given by, respectively
\begin{eqnarray}
g_{H_iZZ}&=&O_{1i}\cos\beta+O_{2i}\sin\beta,\\
g_{H_iH_jZ}&=&(O_{1i}O_{3j}-O_{1j}O_{3i})\sin\beta-(O_{2i}O_{3j}-O_{2j}O_{3i})\cos\beta,
\end{eqnarray}
where $O$ is a 3-by-3 orthogonal matrix, which diagonalizes 
the mass-squared matrix of the neutral Higgs bosons.

According to the recent results of the direct search for the SM Higgs boson at the Fermlab Tevatron, 
the mass range of the SM Higgs boson, $160~{\rm GeV}<m_{h^{\rm SM}}<170$ GeV 
has been excluded at 95\% C.L.~\cite{Phenomena:2009pt}.
Since the upper bound of the lightest Higgs boson mass in the MSSM
is around 135 GeV~\cite{MSSMHiggs_bound}, 
the above excluded mass range is not relevant for the current investigation.

We also consider the constraint on the $\rho$ parameter, which is a measure of the custodial 
$SU(2)$ symmetry breaking.
$\Delta\rho=\Delta_{\tilde{q}}\rho+\Delta_{H}\rho<0.002$ is imposed in our calculation,
where $\Delta_{\tilde{q}}\rho$ and $\Delta_{H}\rho$ are the contributions of stop/sbottom
and the physical Higgs bosons, respectively.
It turns out that as long as we take $m_{{\tilde t}_R}\ll m_{\tilde{q}}$, $\Delta_{\tilde q}\rho$ 
would not exceed the current upper bound even for $m_{{\tilde t}_1}\ll m_{{\tilde b}_1}$ 
due to the suppressed couplings in the loop.
In most of the region under investigation, the mass difference between one of the neutral Higgs bosons
and the charged Higgs bosons is small, which implies the custodial $SU(2)$ symmetry approximately exists, 
we then have $\Delta_H\rho\simeq0$. 

The experimental lower bounds on the masses of the SUSY particles 
are also taken into account, especially
for the lighter stop, chargino ($\chi^\pm_1$), and neutralino ($\chi^0_1$),
which is the lightest supersymmetric particle in our case,  we impose that
$m_{\tilde{t}_1}>95.7$ GeV, $m_{\chi^\pm_1}>94$ GeV and $m_{\chi^0_1}>46$ GeV~\cite{Amsler:2008zzb}.

Currently, the enormous data of the $B$ physics observables  are available, and especially
$B_u\to \tau\nu_\tau$, $\overline{B}\to X_s\gamma$ and 
$B_s\to\mu^+\mu^-$ are relevant for the LHS.
The averaged experimental values of those processes are reported by HFAG~\cite{Barberio:2008fa}:
\begin{eqnarray}
{\rm Br}(B_u\to \tau\nu_\tau)_{\rm exp} &=& 1.41^{+0.43}_{-0.42}\times 10^{-4}, \\
{\rm Br}(\overline{B}\to X_s\gamma)_{\rm exp} &=& (3.52\pm0.23\pm0.09)\times 10^{-4},\\
{\rm Br}(B_s\to\mu^+\mu^-)_{\rm exp} &<& 0.23\times 10^{-7}.
\end{eqnarray}
As mentioned in the Introduction, the first two decay modes can give severe constraints
for the light charged Higgs bosons and the last one could be important for the light neutral Higgs bosons.
However, it is worth to noting that to suppress the $\rho$-parameter corrections, 
the mass difference of the charged Higgs boson and one of the neutral Higgs boson must be small.
Therefore, the above three $B$ decay modes should be simultaneously taken into account. 

At the tree level, the ratio of the two branching ratios of ${\rm Br}(B_u\to \tau\nu_\tau)_{\rm MSSM/SM}$
is given by~\cite{Hou:1992sy}
\begin{eqnarray}
\frac{{\rm Br}(B_u\to \tau\nu_\tau)_{\rm MSSM}}
{{\rm Br}(B_u\to \tau\nu_\tau)_{\rm SM}}
	=\left(1-\tan^2\beta\frac{m^2_{B_u}}{m^2_{H^\pm}}\right)^2\equiv r_H.
\end{eqnarray}
From the latest data~\cite{Amsler:2008zzb}
$|V_{ub}|=(3.95\pm0.35)\times 10^{-3}$, $f_{B}=200\pm20$ MeV,
which are involved in ${\rm Br}(B_u\to \tau\nu_\tau)_{\rm SM}$, it follows that $r_H = 1.28\pm0.52$. 
We will take the 95\% C.L. interval range of $r_H$.
As for ${\rm Br}(\overline{B}\to X_s\gamma)$ and ${\rm Br}(B_s\to\mu^+\mu^-)$, 
we use the public code CPsuperH2.0~\cite{Lee:2007gn} to find the allowed region.

\section{Numerical analysis} \label{sec:numerical}
%
%
\subsection{Light Higgs boson scenario}
\begin{figure}[t]
\includegraphics[width=8cm]{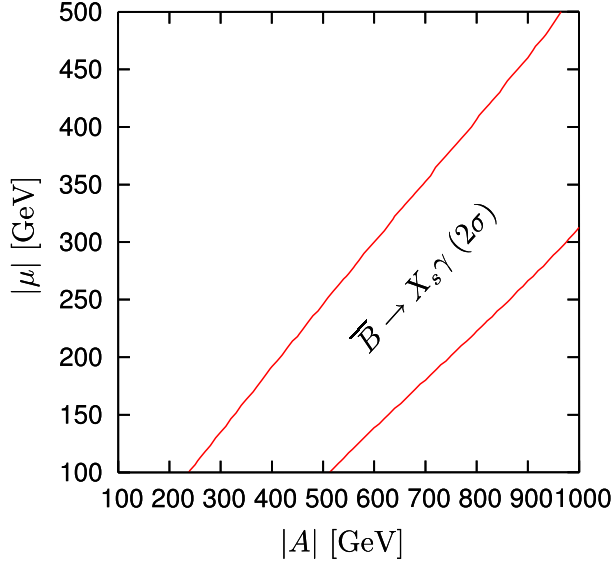}
\caption{The allowed region of $\overline{B}\to X_s\gamma$ in the $|\mu|$-$|A|$ plane. 
$\tan\beta=12$ and $m_{H^\pm}=130$ GeV are taken.}
\label{fig:BSG}
\end{figure}

Before going to show the results of the EWPT, the phenomenologically allowed regions are considered.
The strongest constraint in the LHS mostly comes from $\overline{B}\to X_s\gamma$.
It is well known that the contribution of the charged Higgs boson loop can be cancelled 
by that of the chargino loop. 
To make this mechanism work, the signs of $\mu$ and $A$ are important. 
In Fig.~\ref{fig:BSG}, the 2$\sigma$ allowed region for $\overline{B}\to X_s\gamma$
is shown in the $|\mu|$-$|A|$ plane. 
We take $|A|=|A_t|=|A_b|$, ${\rm Arg}(A_t)={\rm Arg}(A_b)=\pi$, ${\rm Arg}(\mu)=0$,
$\tan\beta=12$, $m_{H^\pm}=130$ GeV, $m_{\tilde{q}}=1200$ GeV, $m_{\tilde{t}_R}=10^{-4}$ GeV, 
$m_{\tilde{b}_R}=1000$ GeV, $M_1=100$ GeV, $M_2=500$ GeV.
The region between the two red lines is allowed. Since the signs of $A$ and $\mu$ are different from 
each other, $X_t$ cannot be vanishing. 
It means that the maximal effect of the stop thermal loop on the strength of the EWPT cannot be realized
in this case. Moreover, the requirement of $100~{\rm GeV}\ltsim |\mu|<|A|$ makes the situation worse.
Here, the lower bound of $|\mu|$ comes from experimental bound of the lighter chargino.
Since the small $X_t$ is favored from the strong first order EWPT point of view, 
we take $\mu=100$ GeV, $A=-300$ GeV in the following discussion.

\begin{figure}[t]
\includegraphics[width=8cm]{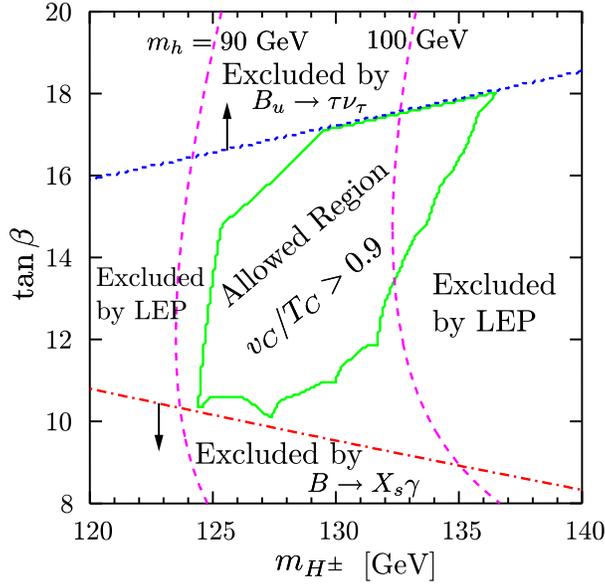}
\caption{The allowed region in the LHS.
We take $A_t=A_b=-300$ GeV, $\mu=100$ GeV, $m_{\tilde{q}}=1200$ GeV, 
$m_{\tilde{t}_R}=10^{-4}$ GeV, and other input parameters are presented in the text.}
\label{fig:EWPT_LHS}
\end{figure}

Figure~\ref{fig:EWPT_LHS} shows the combined results of the LEP and $B$ physics experimental constraints,
and we overlay the $m_h=90,~100$ GeV contours.
The region above the blue dotted line is excluded by the $B_u\to \tau\nu_{\tau}$ mode, and
below the red dashed-dotted line, $B\to X_s\gamma$ exceeds the experimental $2\sigma$ limit.
After taking the LEP constraints into account, 
the region, which is surrounded by the green curves, is phenomenologically allowed.
In this region, the mass of the lightest Higgs boson is mostly between 90 GeV and 100 GeV,
which are shown as the two magenta dashed curves, where the left one stands for $m_h=90$ GeV. 
The mass of the $CP$-odd Higgs boson is approximately given by $m_A\simeq m_h+3~{\rm GeV}$ and that of
the heavy $CP$-even Higgs boson is slightly above the LEP exclusion limit, i.e., 114.4 GeV.
For this parameter set, the constraint of $B_s\to\mu^+\mu^-$ 
does not play a role unless $\tan\beta\gtsim 20$.
It is found that $v_C/T_C$ in the allowed region is larger than 0.9, and the maximal value of it is found to be
\begin{eqnarray}
\frac{v_C}{T_C}=\frac{107.10~{\rm GeV}}{116.27~{\rm GeV}}=0.92
\label{max_vcTc}
\end{eqnarray}
for $\tan\beta=10.1$ and $m_{H^\pm}=127.4$ GeV. 
This value is far below the required sphaleron decoupling condition as shown in Table~\ref{tab:sph_cond}.

\begin{figure}[t]
\includegraphics[width=7cm]{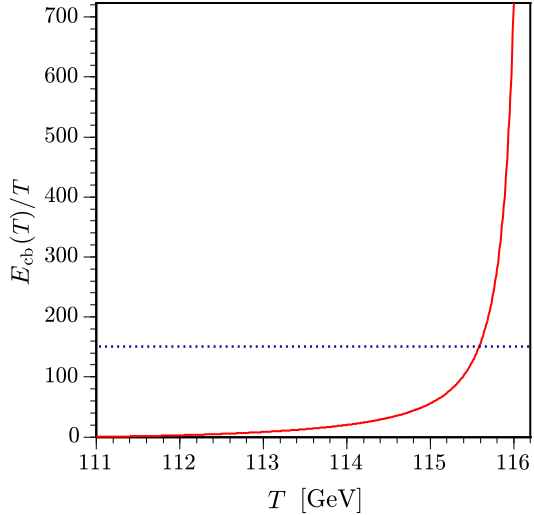}
\caption{$E_{\rm cb}(T)/T$ as a function of $T$ in the LHS. 
The blue dotted line shows $E_{\rm cb}(T_N)/T_N=150.386$, where $T_N=115.59$ GeV.
We take $\tan\beta=10.1$ and $m_{H^\pm}=127.4$ GeV 
and other parameters are the same as in the Fig.~\ref{fig:EWPT_LHS}}
\label{fig:Ecb_LHS}
\end{figure}

\begin{figure}[t]
\includegraphics[width=7.5cm]{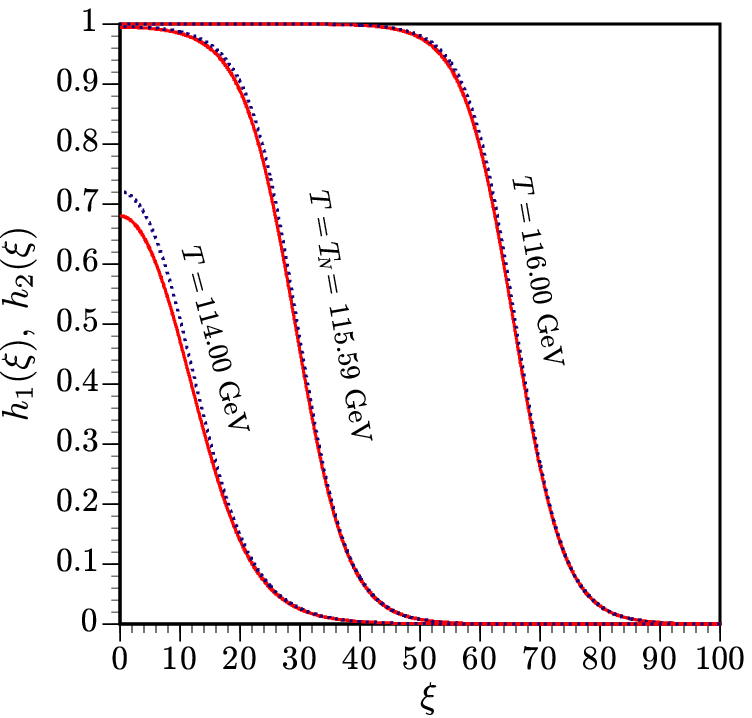}
\hspace{0.5cm}
\includegraphics[width=8cm]{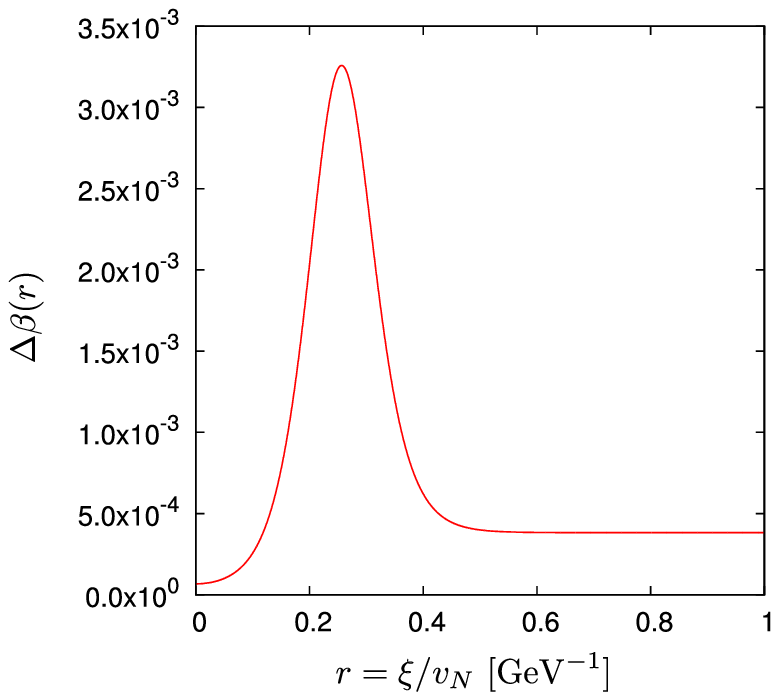}
\caption{Left: $h_{1}(\xi)$ (red solid curve) and $h_2(\xi)$ (blue dotted curve) 
are plotted for $T=116.00, 115.59~(=T_N), 114.00$ GeV.
Right: $\Delta\beta(r)$ as function of $r=\xi/v_N$.
The input parameters are the same as in Fig.~\ref{fig:Ecb_LHS}}.
\label{fig:bubbles_LHS}
\end{figure}

As we discussed in Sec.~\ref{sec:cri_bub}, the temperature at which the EWPT occurs
is not $T_C$ but somewhat below it. 
With the same input parameter set as is used in Eq.~(\ref{max_vcTc}),
we find the critical bubble solutions and evaluate their nucleation temperature.
In Fig.~\ref{fig:Ecb_LHS}, $E_{\rm cb}(T)/T$ is shown as a function of $T$.
The blue dotted line corresponds to $E_{\rm cb}(T_N)/T_N=150.39$, where $T_N=115.59$ GeV.
Since $(T_C-T_N)/T_C=5.8\times 10^{-3}$, the supercooling is small.
The left plot in Fig.~\ref{fig:bubbles_LHS} shows that the bubble walls 
at $T=116.00, 115.59~(=T_N), 114.00$ GeV. The red solid curves represent $h_1(\xi)$, and 
the blue dotted curves correspond to $h_2(\xi)$.
As $T$ decreases, the energy of the broken vacuum becomes lower, correspondingly, 
and the radius of the bubble becomes smaller as it should be.
It is found that the wall width is rather thick, which is consistent 
with the previous studies~\cite{Moreno:1998bq,cpv_bubble2}.
In the standard mechanism of electroweak baryogenesis in the thick wall regime, 
the variation of $\tan\beta(r)$ along 
the line connecting the broken phase and symmetric phase is crucial to the amount of the net BAU.
Here, we define
\begin{eqnarray}
\Delta\beta(r) = \beta(r)-\beta(T),
\end{eqnarray}
where $\tan\beta(T)=v_u(T)/v_d(T)$ 
and $\tan\beta(r)=\rho_u(r)/\rho_d(r)=\tan\beta(T)h_2(\xi)/h_1(\xi)$.
$\Delta\beta(r)$ at $T_N$ is shown in the right plot in Fig.~\ref{fig:bubbles_LHS}.
This shows that the behaviors of $\rho_d(r)$ and $\rho_u(r)$ in the intermediate of $r$ 
are slightly different. We observe that $\rho_d(r)$ approaches to zero somewhat faster 
than $\rho_u(r)$ does as $r$ increases.
Such an enhancement might play a role in generating the BAU.

At $T_N$, $v_N/T_N$ is given by
\begin{eqnarray}
\frac{v_N}{T_N}=\frac{116.73~{\rm GeV}}{115.59~{\rm GeV}}=1.01.
\end{eqnarray}
One can see that there is the $\mathcal{O}(10)$\% enhancement in $v/T$ compared to that at $T_C$.
However, as we discussed in Sec.~\ref{sec:sph_dec} the sphaleron decoupling condition at $T_N$ 
is given by
\begin{eqnarray}
\frac{v_N}{T_N}>1.38.
\end{eqnarray}
Therefore, the sphaleron process is not decoupled at $T_N$ either in this parameter region.
We search for the maximal value of $v_C/T_C$ scanning $\tan\beta$ and $m_{H^\pm}$
for $m_{\tilde{q}}=1300,~1400$ and $1500$ GeV.
Our findings in the LHS are summarized in Table~\ref{tab:summary_LHS}.
Every case is more or less the same as the first one discussed here.
We thus conclude that there is no region where the sphaleron decoupling condition is satisfied in the LHS.

\begin{table}[t]
\caption{Examples of the first order EWPT in the LHS. 
$|A_b|=|A_t|=300$ GeV,~$m_{\tilde{t}_R}=10^{-4}$ GeV,~ 
$m_{\tilde{b}_R}=1000$ GeV,~$|\mu|=100$ GeV,~$M_1=100$ GeV,~$M_2=500$ GeV,
~${\rm Arg}(A_t)={\rm Arg}(A_b)=\pi$,~${\rm Arg}(\mu)=0$.}
\begin{center}
\begin{tabular}{|c||c|c|c|c|}
\hline
$m_{\tilde{q}}$ (GeV) & 1200 & 1300 & 1400 & 1500 \\
\hline\hline
$m_{\tilde{t}_1}$ (GeV) & 164.76 & 165.66 & 166.37 & 166.95 \\
%
$m_{\chi^\pm_1}$ (GeV) & 94.86 & 94.80 & 94.77 & 94.72\\
%
$m_{\chi^0_1}$ (GeV) & 65.14 & 65.04 & 64.99 & 64.92\\
$m_{\chi^0_2}$ (GeV) & 108.31 & 108.26 & 108.24 & 108.21\\
$m_{\chi^0_3}$ (GeV) & 129.46 & 129.50 & 129.52 & 129.56\\
\hline
$\tan\beta$ & 10.11 & 9.87 & 9.75 & 9.57 \\
$m_{H^\pm}$ (GeV) & 127.40 & 127.40 & 127.50 & 127.50 \\ 
$m_{H_1}$ (GeV) & 94.04 & 93.95 & 94.03 & 93.97 \\
$m_{H_2}$ (GeV) & 97.82 & 97.85 & 97.98& 97.99 \\
$m_{H_3}$ (GeV) & 116.47 & 117.13 & 117.72 & 118.31 \\
$g^2_{H_1VV}$ & 0.228 & 0.228 & 0.227 & 0.227 \\
$g^2_{H_2VV}$ & 0.000 & 0.000 & 0.000 & 0.000 \\
$g^2_{H_3VV}$ & 0.772 & 0.772 & 0.773 & 0.773 \\
\hline\hline
 & & & & \\[-0.4cm]
	$v_C/T_C$ 
	& $\displaystyle{\frac{107.096}{116.274}=0.921}$ 
	& $\displaystyle{\frac{107.512}{116.496}=0.923}$ 
	& $\displaystyle{\frac{107.769}{116.770}=0.923}$
	& $\displaystyle{\frac{107.915}{117.045}=0.922}$ \\
$\tan\beta_C$ & 13.803 & 13.640 & 13.597 & 13.455 \\
	$v_N/T_N$ 
	& $\displaystyle{\frac{116.727}{115.585}=1.010}$ 
	& $\displaystyle{\frac{ 117.155}{115.798}=1.012}$ 
	& $\displaystyle{\frac{117.404}{116.067}=1.012}$
	& $\displaystyle{\frac{117.531}{116.339}=1.010}$ \\
$\tan\beta_N$ & 13.676 & 13.503 & 13.453 & 13.307 \\
%
%
$E_{\rm cb}(T_N)/T_N$ & 150.386 & 150.379 & 150.370 & 150.360 \\
\hline\hline
$\mathcal{E}$ & $1.769$ & $1.770$ & $1.770$ & $1.771$ \\
$\mathcal{N}_{\rm tr}$ & $6.652$ & $6.658$ & $6.662$ & $6.667$ \\
$\mathcal{N}_{\rm rot}$ & $12.266$& $12.253$ & $12.240$ & $12.229$ \\
$v_N/T_N>$ & $1.383$ & $1.382$ & $1.382$ & $1.380$ \\
\hline
\end{tabular}
\end{center}
\label{tab:summary_LHS}
\end{table}

\subsection{The decoupling limit}
In the decoupling limit, $h$ is the SM-like Higgs boson and must be heavier than 114.4 GeV.
The constraints from $B$ physics are not so strong in this region.
We then consider the no-mixing scenario, $X_t=0$ and $A_b=A_t$ are taken.
Other parameters are the same as those in the LHS discussed in the previous subsection.
For $m_{\tilde{q}}\ltsim$ 1600 GeV, there is no allowed region such that $m_h>114.4$ GeV.
As an example, we take $m_{\tilde{q}}=1700$ GeV and scan $\tan\beta$ and $m_{H^\pm}$
in the ranges: $1\le\tan\beta\le50$, $100~{\rm GeV}\le m_{H^\pm}\le 1000~{\rm GeV}$.
The maximal value of $v_C/T_C$ is realized in the case of 
$\tan\beta=42.6$ and $m_{H^\pm}=1000$ GeV.
$v/T$at $T_C$ and $T_N$ are found to be
\begin{eqnarray}
\frac{v_C}{T_C}=\frac{114.46~{\rm GeV}}{116.99~{\rm GeV}}=0.95,\quad
\frac{v_N}{T_N}=\frac{121.45~{\rm GeV}}{116.22~{\rm GeV}}=1.05.
\end{eqnarray}
Similar to the LHS, $v/T$ is ameliorated at $T_N$ by about 10\%.
However, it is still not strong enough to satisfy the sphaleron decoupling condition which is given by
$v_N/T_N>1.38$.

We search for the allowed region varying $m_{\tilde{q}}=1800,~1900,~2000$ GeV 
and investigate $v_C/T_C$ and $v_N/T_N$.
The results are summarized in Table \ref{tab:summary_dec}.
There is no significant change between the first case discussed above and the others.
We also scan in the region, $m_{\tilde{q}}>2000$ GeV. 
The larger $m_{\tilde{q}}$ allows the smaller $\tan\beta$ region,
which implies that the top Yukawa coupling becomes stronger.
On the other hand, $m_h$ also becomes larger through the one-loop top/stop corrections. 
The numerical results show that the first order EWPT is getting weaker and weaker
as $m_{\tilde{q}}$ increases.

\begin{table}[t]
\caption{Examples of the first order EWPT in the no-mixing scenario.
$|A_t|=|A_b|=|\mu|/\tan\beta$, $m_{\tilde{t}_R}=10^{-4}$ GeV,~ 
$m_{\tilde{b}_R}=1000$ GeV,~$|\mu|=100$ GeV,~$M_1=100$ GeV,~$M_2=500$ GeV,
~${\rm Arg}(A_t)={\rm Arg}(A_b)={\rm Arg}(\mu)=0$.}
\begin{center}
\begin{tabular}{|c||c|c|c|c|}
\hline
$m_{\tilde{q}}$ (GeV) & 1700 & 1800 & 1900 & 2000 \\
\hline\hline
$m_{\tilde{t}_1}$ (GeV) & 170.73 & 170.76 & 170.78 & 170.81 \\
%
$m_{\chi^\pm_1}$ (GeV) & 96.80 & 95.69 & 95.06 & 94.66\\
%
$m_{\chi^0_1}$ (GeV) & 68.25 & 66.46 & 65.45 & 64.82 \\
$m_{\chi^0_2}$ (GeV) & 109.73 & 108.92 & 108.45 & 108.16 \\
$m_{\chi^0_3}$ (GeV) & 128.15 & 128.91 & 129.34 & 129.59 \\

\hline
$\tan\beta$ & 42.62 & 15.10 & 10.97 & 9.35 \\
$m_{H^\pm}$ (GeV) & 1000.00 & 1000.00 & 1000.00 & 1000.00 \\ 
$m_{H_1}$ (GeV) & 114.40 & 114.42 & 114.40 & 114.48 \\
$m_{H_2}$ (GeV) & 994.09 & 996.45 & 996.61 & 996.66 \\
$m_{H_3}$ (GeV) & 994.11 & 996.54 & 996.77 & 996.88 \\
$g^2_{H_1VV}$ & 1.000 & 1.000 & 1.000 & 1.000 \\
$g^2_{H_2VV}$ & 0.000 & 0.000 & 0.000 & 0.000 \\
$g^2_{H_3VV}$ & 0.000 & 0.000 & 0.000 & 0.000 \\
\hline\hline
 & & & & \\[-0.4cm]
	$v_C/T_C$ 
	& $\displaystyle{\frac{111.461}{116.993}=0.953}$ 
	& $\displaystyle{\frac{111.460}{117.007}=0.953}$ 
	& $\displaystyle{\frac{111.483}{116.994}=0.953}$
	& $\displaystyle{\frac{111.440}{117.060}=0.952}$ \\
$\tan\beta_C$ & 42.966 & 15.171 & 11.022 & 9.394 \\
	$v_N/T_N$ 
	& $\displaystyle{\frac{121.454}{116.221}=1.045}$ 
	& $\displaystyle{\frac{121.452}{116.236}=1.045}$ 
	& $\displaystyle{\frac{121.478}{116.222}=1.045}$
	& $\displaystyle{\frac{121.424}{116.288}=1.044}$ \\
$\tan\beta_N$ & 42.955 & 15.168 & 11.019 & 9.392 \\
%
%
$E_{\rm cb}(T_N)/T_N$ & 150.366 & 150.370 & 150.364 & 150.360 \\
\hline\hline
$\mathcal{E}$ & $1.773$ & $1.773$ & $1.773$ & $1.773$ \\
$\mathcal{N}_{\rm tr}$ & $ 6.677$ & $6.677$ & $6.678$ & $6.678$ \\
$\mathcal{N}_{\rm rot}$ & $12.211$& $12.210$ & $12.210$ & $12.209$ \\
$v_N/T_N>$ & $1.379$ & $1.379$ & $1.379$ & $ 1.379$ \\
\hline
\end{tabular}
\end{center}
\label{tab:summary_dec}
\end{table}

\section{Conclusion and discussions} \label{sec:con_dis}
We have reanalyzed the strong first order EWPT in the MSSM 
considering the experimental results such as the LEP and the latest $B$ physics data.
$v/T$ was evaluated at both $T_C$ and $T_N$ in the LHS and ordinary decoupling limit.
In the LHS with a light stop, the no-mixing scenario, which can maximize the stop thermal loop effect,
cannot be realized due to the severe constraint from $\overline{B}\to X_s\gamma$. 
Combining with the other experimental data, especially the LEP 
and $B_u\to \tau\nu_\tau$ mode, the allowed region becomes more constrained.

The sphaleron decoupling condition was also calculated 
using the one-loop effective potential at finite temperature.
We found that the contributions of the zero mode factors coming from 
the fluctuations around sphaleron can be as large as about 10\%.
In the phenomenologically allowed region, 
the typical values of the sphaleron decoupling conditions at $T_N$ are found to be $v_N/T_N\gtsim 1.38$.

It is observed that about 10\% enhancement of $v/T$ is possible at $T_N$ in comparison with that at $T_C$.
We find that $v_N/T_N\simeq 1.01$ in the LHS and $v_N/T_N\simeq 1.05$ in the decoupling limit.
However, such a strength of the EWPT is not enough to satisfy the sphaleron decoupling condition.

Although electroweak baryogenesis in the MSSM seems to be infeasible, 
it may be possible to circumvent the above negative conclusion. 
Some comments on the possible loopholes are in order.
Firstly, the sphaleron decoupling condition is supposed to be imposed at the temperature at which 
the EWPT ends, which can relax the decoupling condition.
Secondly, our calculations, which are based on the one-loop effective potential, can be
modified by the higher-order loop contributions such as the two-loop effects~\cite{2loope_ewpt}.
Lastly, as we mentioned briefly in Sec.~\ref{sec:EWPT}
the scalar potential can be extended in such a way that the lighter stop also has
the VEV, which can develop the CCB vacuum. 
In Ref.~\cite{Carena:2008vj} it is claimed that the EWPT is strong first order when the electroweak
vacuum is metastable and the CCB vacuum is the global minimum.
Since the time scale of the decay of the metastable vacuum is longer than the age of the Universe,
this scenario is viable.
The two-loop contributions are also included in their calculation.
It is interesting to investigate whether or not the EWPT is still strong enough to satisfy
the sphaleron decoupling conditions that we have considered in this article.
A devoted study of the EWPT and the sphaleron decoupling condition
based on the two-loop effective potential will be given elsewhere~\cite{FS}.

\appendix
%
%
\section{The mass-squared matrices of stop and sbottom}\label{app:mass_squared}
Here, we give the expressions of the mass-squared matrices of the stop and sbottom
\begin{eqnarray}
\mathcal{M}^2_{\tilde{t}}&=&
	\left(
		\begin{array}{cc}
		m^2_{\tilde{q}}+\frac{|y_t|^2}{2}v^2_u+\frac{1}{8}\bigg(g^2_2-\frac{g^2_1}{3}\bigg)
		(v^2_d-v^2_u) &
		\frac{y^*_t}{\sqrt{2}}(A^*_te^{-i\vartheta}v_u-\mu v_d) \\
		\frac{y_t}{\sqrt{2}}(A_te^{i\vartheta}v_u-\mu^* v_d) &
		m^2_{\tilde{t}_R}+\frac{|y_t|^2}{2}v^2_u+\frac{g^2_1}{6}(v^2_d-v^2_u)
		\end{array}
	\right), \\
\mathcal{M}^2_{\tilde{b}}&=&
	\left(
		\begin{array}{cc}
		m^2_{\tilde{q}}+\frac{|y_b|^2}{2}v^2_d-\frac{1}{8}\bigg(g^2_2+\frac{g^2_1}{3}\bigg)
		(v^2_d-v^2_u) &
		\frac{y^*_b}{\sqrt{2}}(A^*_bv_d-\mu e^{i\vartheta} v_u) \\
		\frac{y_b}{\sqrt{2}}(A_bv_d-\mu^* e^{-i\vartheta} v_u) &
		m^2_{\tilde{b}_R}+\frac{|y_b|^2}{2}v^2_d-\frac{g^2_1}{12}(v^2_d-v^2_u)
		\end{array}
	\right),
\end{eqnarray}
where $y_t$ and $y_b$ are the top and bottom Yukawa couplings, respectively.

\begin{acknowledgments}
E.S. would like to thank Y. Okada and J.-S. Lee for useful discussions.
E.S was supported in part by the National Science Council of Taiwan, R.O.C. under
Grant No. NSC 97-2112-M-008-002-MY3.
\end{acknowledgments}


\end{document}